\documentclass[epj]{svjour}

\usepackage{graphics}
\usepackage{epsfig}

\journalname{The European Physical Journal E}

\newcommand{\be}{\begin{equation}}
\newcommand{\ee}{\end{equation}}
\newcommand{\bea}{\begin{eqnarray}}
\newcommand{\eea}{\end{eqnarray}}
\newcommand{\bfig}{\begin{figure}}
\newcommand{\efig}{\end{figure}}
\newcommand{\bfet}{\begin{figure*}}
\newcommand{\efet}{\end{figure*}}
\newcommand{\bc}{\begin{center}}
\newcommand{\ec}{\end{center}}
\newcommand{\beqa}{\begin{eqnarray}}
\newcommand{\eeqa}{\end{eqnarray}}
\newcommand{\bay}{\begin{array}}
\newcommand{\eay}{\end{array}}
\newcommand{\btab}{\begin{tabular}}
\newcommand{\etab}{\end{tabular}}

\newcommand{\szz}{\sigma_{zz}}

\newcommand{\suu}{\sigma_{11}}
\newcommand{\sdd}{\sigma_{22}}
\newcommand{\sud}{\sigma_{12}}

\newcommand{\uuu}{u_{11}}
\newcommand{\udd}{u_{22}}
\newcommand{\uud}{u_{12}}

\begin{document}

\title{
From the stress response function (back) to the sandpile `dip'}

\author{
A.P.F. Atman$^\clubsuit$,
P. Brunet$^\clubsuit$,
J. Geng$^\spadesuit$,
G. Reydellet$^\clubsuit$,
P. Claudin$^\clubsuit$,
R.P. Behringer$^\spadesuit$ and
E. Cl\'ement$^\clubsuit$.}

\institute{
$\clubsuit$
Laboratoire de Physique et M\'ecanique des Milieux H\'et\'erog\`enes,\\
ESPCI, 10 rue Vauquelin, 75231 Paris Cedex 05, France.\\
$\spadesuit$
Department of Physics \& Center for Nonlinear and Complex Systems,\\
Duke University, Durham NC, 27708-0305, USA.}

\date{\today}

\abstract{
We relate the pressure `dip' observed at the bottom of a sandpile prepared
by successive avalanches to the stress profile obtained on sheared granular
layers in response to a localized vertical overload. We show that, within a
simple anisotropic elastic analysis, the skewness and the tilt of the response
profile caused by shearing provide a qualitative agreement with the
sandpile dip effect. We conclude that the texture anisotropy produced by the
avalanches is in essence similar to that induced by a simple shearing --
albeit tilted by the angle of repose of the pile. This work also shows that
this response function technique could be very well adapted to probe the
texture of static granular packing.
}


\authorrunning{A.P.F. Atman \emph{et al.}}
\titlerunning{From the stress response function (back) to the sandpile `dip'}

\maketitle

%_____________________________________________________________________________
The stress distribution below a pile of sand has been one of the
problematic issues of the statics of granular materials in physics over the
last few years \cite{reviews}. In fact, experiments have shown that, when a
granular pile is prepared from a point source, the bottom pressure profile
has a clear local minimum -- a `dip' -- below the apex
\cite{SN81,BHB97,VHCBC99}. The existence of this
pressure dip has been strongly debated, and it is now well established that
the presence or absence of this dip is closely related to the preparation
history of the pile. This was demonstrated by Vanel \emph{et al.}
\cite{VHCBC99}. Using the same sand and experimental apparatus, these authors
could generate the stress dip using a localized deposition technique or
cause the dip to disappear by constructing a pile in successive horizontal
layers. Similar conclusions were reached for two dimensional heaps with
photo-elastic grains \cite{GLBH01}, and in numerical simulations
\cite{LCH92,L97,M98}.

This interesting effect has inspired the development of new models to
describe how forces are transmitted in dense granular materials. Among
them are those proposed by Bouchaud \emph{et al.}, initially developed
in the context of the sand pile dip \cite{BCC95,WCCB96,WCC97,C99}, and
further extended to other geometries like that of the silo
\cite{C99,VCBCC00}. This approach is also intended to describe a
collection of systems including dense colloids, granular matter or
foams \cite{CWBC98}. At the macroscopic level, these features are
modelled by hyperbolic, partial differential equations (PDE) for the
stress tensor. Although no explicit link was established, the
characteristics of these hyperbolic equations were intuitively thought
to be related to the mesoscopic `force chain' network whose structure
and orientation were shaped by the previous history of the granular
assembly -- see also \cite{YY,B04} concerning force chains. Plasticity
theories for granular deformations are also of hyperbolic type,
although conceptually different than the previous cited models.
From the classical, soil mechanics point of view, below
the plastic threshold, granular material is thought to behave as an
effective elastic material with PDE's that belong to the elliptic
class \cite{W90}. Finally, sound wave propagation techniques and
numerical simulations of confined granular assemblies indicate that
that assesment of effective elastic constitutive relations is still an
open and difficult issue \cite{MGJS99}.

\bfet[t]
\bc
\epsfxsize=0.25\linewidth
\epsfbox{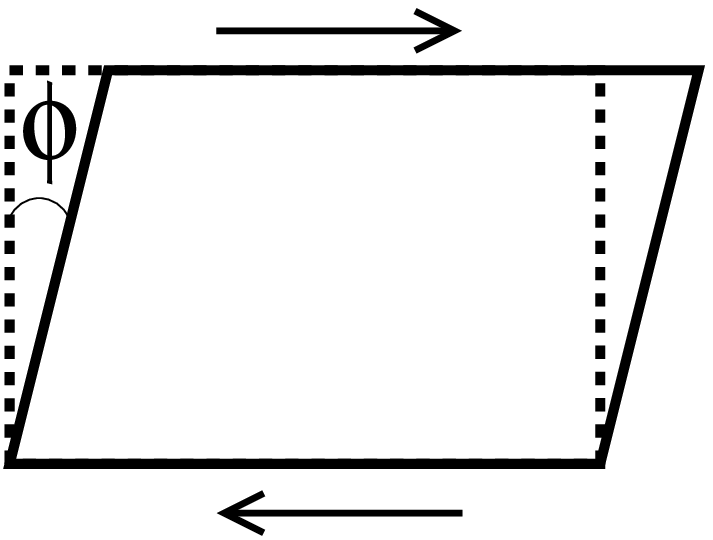}
\hfill
\epsfxsize=0.4\linewidth
\epsfbox{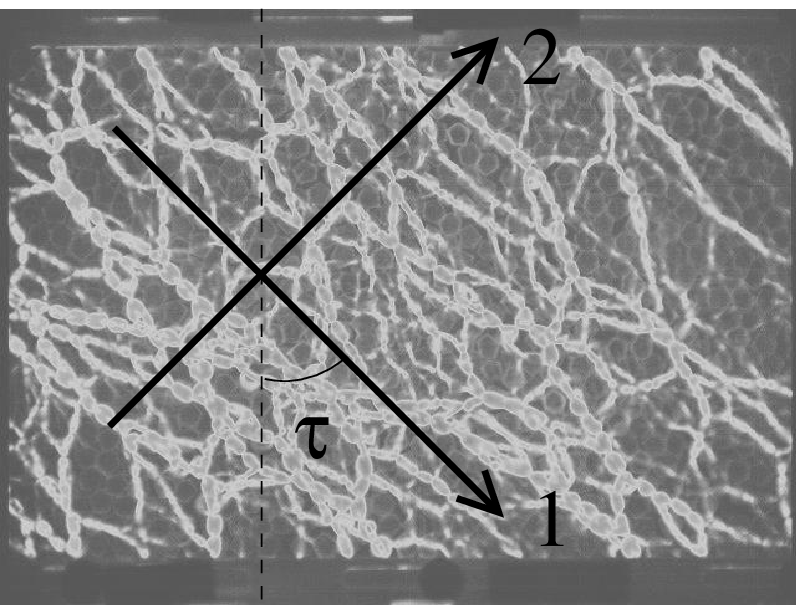}
\hfill
\epsfxsize=0.3\linewidth
\epsfbox{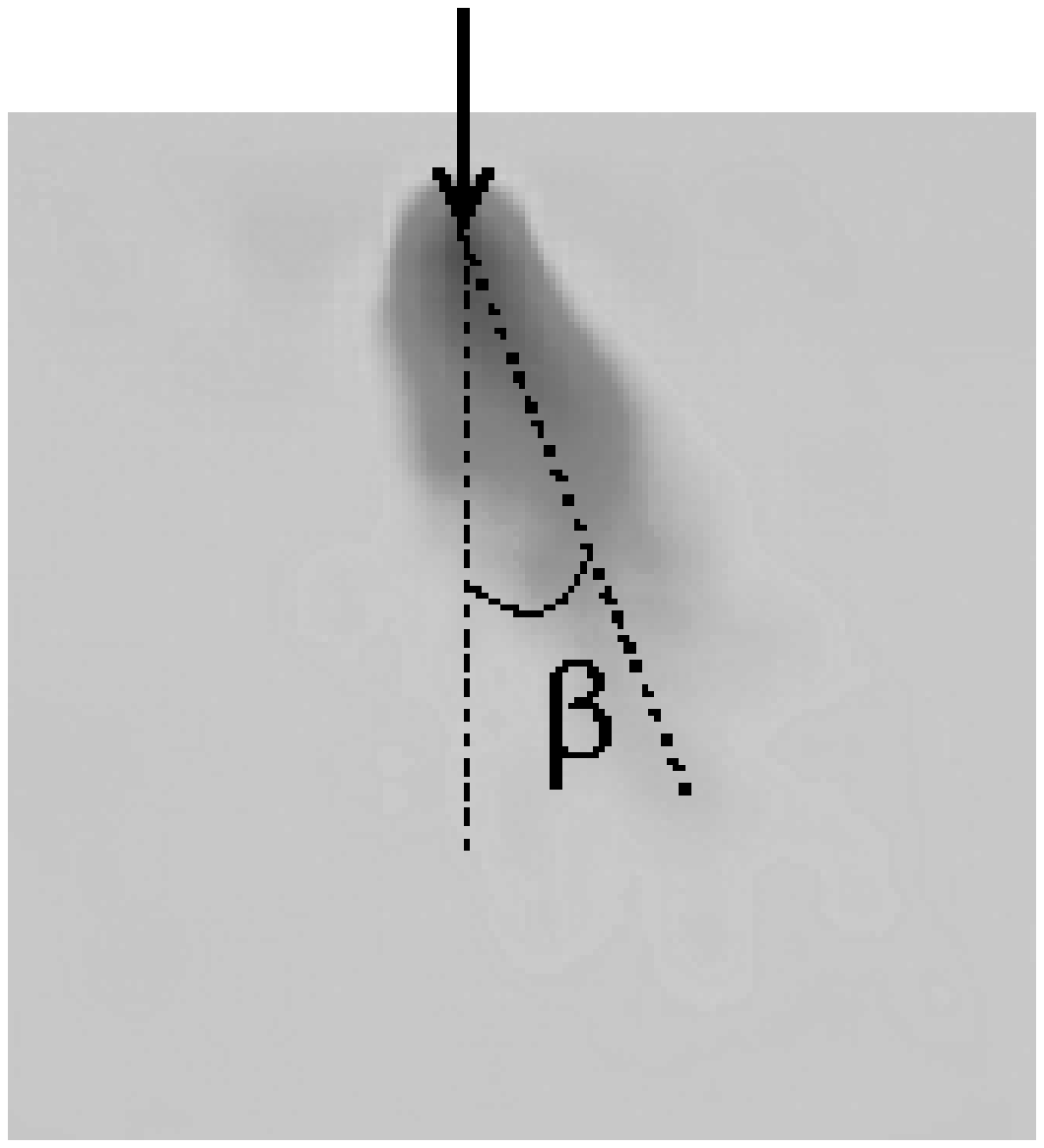}
\ec
\caption{Left: Sketch of the 2D shear box. The system is strained up to an
angle $\phi$. Middle: Visualization of the force chains after shearing.
The analysis of their orientations shows that a direction at $\tau=45^\circ$
is prefered \cite{GRCB03}.
Right: Average response to a vertical overlaod after shearing with a strain
angle $\phi=5^\circ$. The stress maximum is oriented at $\beta=22^\circ$ to
the vertical.
\label{fig:exp2D}}
\efet

In order to distinguish between the very different mathematics of
hyperbolic and elliptic PDEs a stress response experiment was proposed
\cite{BCC95,dG98}. For instance, the pressure profile measured at the
bottom of an isotropic elastic horizontal layer in response to a
localized vertical overload at its top surface should be a single
broad peak, while hyperbolic models would yield in 2D two thin peaks
or a ring in 3D. Several experiments
\cite{SR00,RC01,SRCCL01,GHLBRV01,MJN02,GRCB03,SMJN03} and simulations
\cite{EC97,M00,BCCZ02,GG02a,GG04,SVG02,OP04,ABGRCCBC04,GWM04,GG05}
have recently addressed the issue of the stress response in a granular
layer.  Collectively, they have demonstrated two key points:\\ (i) the
shape of the pressure response profiles is generally not in agreement
with the predictions of the hyperbolic models. For the generic case of
a disordered packing of frictional grains, the measured profiles show
an elliptic-like behavior with one single peak broadening in
proportion to the thickness/depth of the layer -- except in
\cite{SR00}. For well ordered packing, the measured response functions
may produce rays diverging from the source, but this type of response
is also compatible with anisotropic elasticity \cite{GG02a,GG02,OBCS03};\\
(ii) the geometry of the response function is in fact simpler than
that of the pile, and offers rich possibilities. Response function
measurements directly detect any symmetry breaking due to texture
originating from either a specific preparation or from any other
external action on the pile. The layer can be loose, dense, compacted,
sheared, avalanched, sedimented, ordered, vibrated, and so on.\\
Nevertheless, there are still a number of unresolved questions. For
instance in the limiting case of isostatic pilings, numerical
simulations \cite{HTW01,KN04} and theoretical arguments
\cite{EG99,TW99,BB02,M02,B04} indicate that a hyperbolic equations should
describe the stress propagation, although this is not in agreement
with the work of Roux \cite{R00}.

Experiments have shown that the response to a point force is very
sensitive to the preparation of the system \cite{ABGRCCBC04}. For
instance, for a 2D system which has been subjected to strain in a 2D
shear box, the response function is skewed in the direction of the
shear \cite{GRCB03} showing a strain-induced anisotropy. Here, we
extend this result to 3D granular assemblies and propose a relation to
the pressure `dip' observed at the bottom of a sand pile when prepared
by successive avalanches. Interestingly, anisotropy induced by preparation
was suggested by Savage \cite{S98} to explain this phenomenon. Note
that the occurence of a stress solution with a dip can also be produced
in a model pile composed of an elastic core and plastic wings
\cite{DCG00}.

The paper is organized as follows. We first describe experimental
results obtained for the response of a sheared granular layer in 2D
and in 3D (section \ref{secI}) and for a layer prepared by successive
avalanches. We then present a theoretical anisotropic elastic analysis
of these experimental data in 2D and a numerical analysis in 3D
(section \ref{secII}).  Thereafter, we numerically solve the case of a
conical heap (section \ref{secIII}).  We close with conclusions and
suggested perspectives.

%_____________________________________________________________________________
\section{Experimental response function on sheared granular layers}
\label{secI}

The response function experiments of interest here, were carried out
in shear cell geometries in two and three dimensions. In both cases,
the cell consisted of vertical boundaries that could be tilted
quasi-statically by an angle $\phi$ with respect to the vertical
axis. This process deformed the samples from an original rectangular
geometry to that of a parallelogram.

In 2D, the spacing between the horizontal boundaries was maintained
strictly constant, so that the sample volume also remained
constant. The particles were pentagons made of a photo-elastic
material, which allows a direct measurement of the local force
response and of the force chains. More details on these experiments
can be found in Geng \emph{et al.} \cite{GRCB03}.  As we can see from
figure \ref{fig:exp2D}, the salient features of these series of
experiments are (i) the force chain network is oriented at
$45^{\circ}$ from the horizontal axis, which is the principal
compressive direction and (ii) the response to a vertical force is
tilted with an angle $\beta$ with respect of the vertical direction,
which illustrates perfectly the symmetry breaking due to shearing. In
a previous contribution we argue that this $\tau=45^\circ$ angle is
simply related to the principal axes of compression and dilation
(respectively directions $1$ and $2$ on figure \ref{fig:exp2D}).

\bfet[t]
\bc
\epsfxsize=0.4\linewidth
\epsfbox{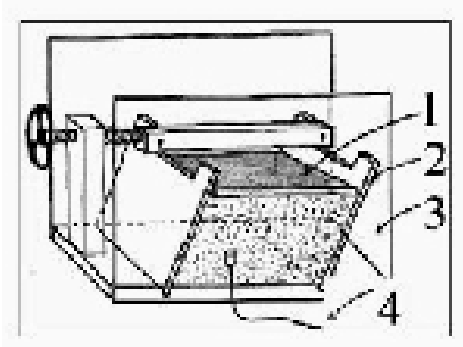}
\hfill
\epsfxsize=0.57\linewidth
\epsfbox{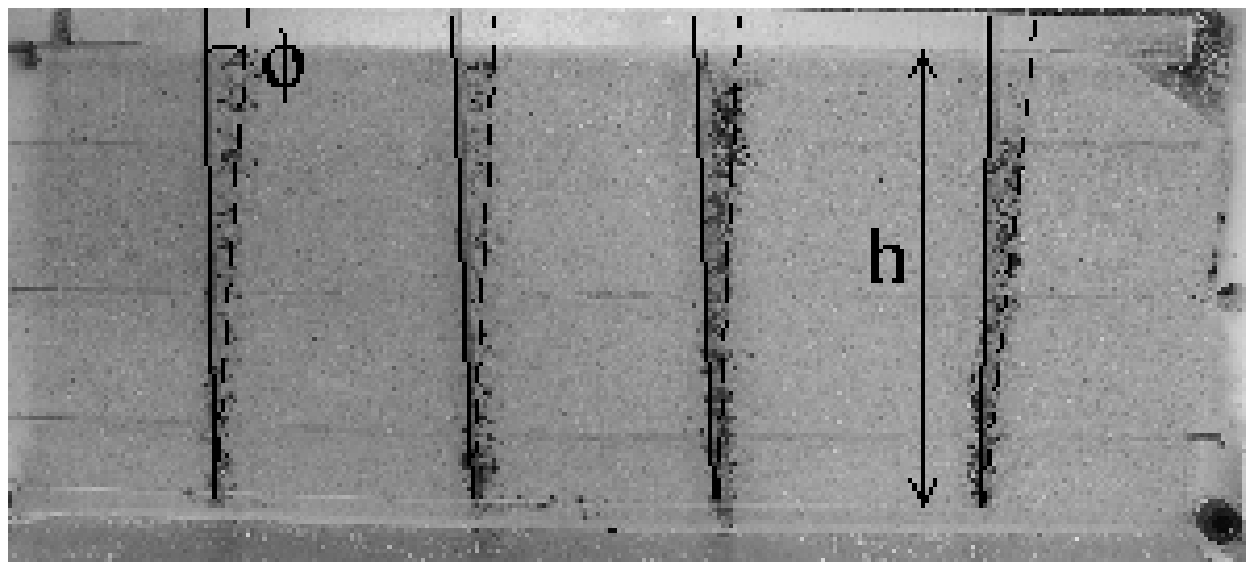}
\ec
\caption{Left: Sketch of the 3D shear box. $1$ top lid, $2$ tiltable lateral
boundaries, $3$ fix glass wall, $4$ capacitive stress probe.
Right: Visualization of the grain displacement due to shear. Coloured grains
have been put along four vertical lines before shearing. These grains are
located along inclined lines afterwards, so that the strain field of the layer
looks reasonnably homogeneous.
\label{fig:exp3D}}
\efet

In 3D, we carry out two experiments to extend the study of the
influence of shear on the texture of a granular assembly. First, we
built an apparatus similar to the shear cell already used in
2D. Second, we deposited the material in a horizontal layer by the
superposition of successive avalanches.

\emph{The shear cell} -- The shear cell is sketched in the left part
of figure \ref{fig:exp3D}.  In order to probe the mechanical properties
and the symmetries associated with the induced structure, we mounted a
capacitive stress probe at the bottom of the cell (labeled $4$ in
figure \ref{fig:exp3D}); the applied force on the top (labeled $1$)
is moved horizontally. Specifically, the force response measurements
use a low frequency modulation of the localized stress imposed at the
top of the sample, and lock-in detection of the vertical pressure
response on the bottom, as described by Reydellet \emph{et al.}
\cite{RC01}. The lock-in detection provided a large signal/noise
ratio, which allowed us to apply tiny forces, of the same order as the
weight of few grains. This prevented any significant deformation of
the packing which would have modified the fragile structure of
contacts. A horizontal layer of grains (Fontainebleau sand, slightly
polydisperse round grains, $d \sim 300 \mu$m size) was initially
prepared by pouring the material into shear box of horizontal size
$l=25$cm. The lateral width of the box was $w=20$cm. The total height
$h$ of sand ranged between $4$ and $15$cm.

\bfig[b]
\bc
\epsfxsize=\linewidth
\epsfbox{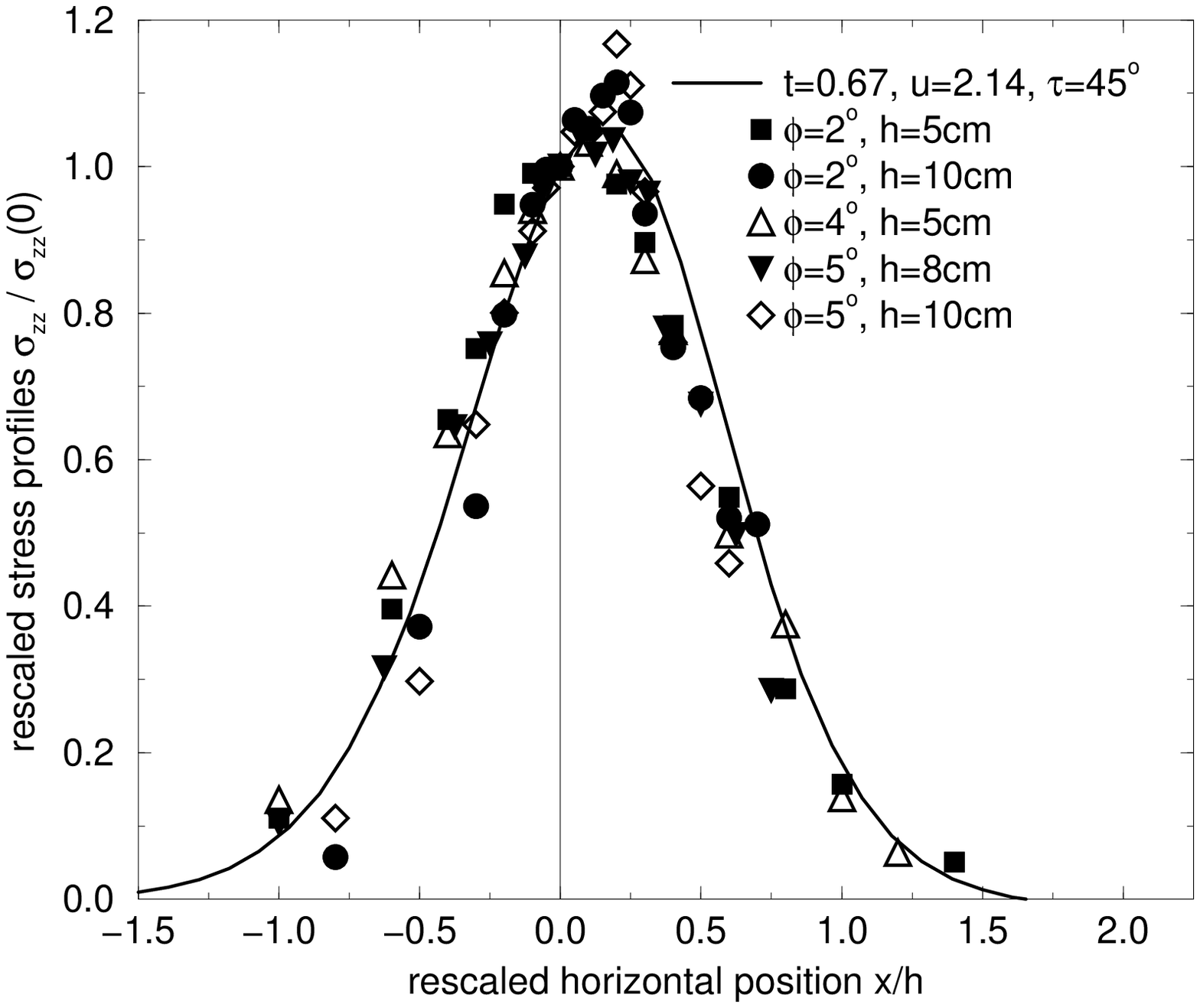}
\ec
\caption{Response profiles of a 3D sheared layer of Fontainebleau sand for
various values of the layer thickness $h$ and the strain angle $\phi$.
The maxima of the profiles correspond to a tilt angle $\beta$ of
$8^\circ \pm 1^\circ$. The collapse of the data is resonnably fitted by
the \textsc{sem} (see section \ref{secII} for the definition of
the \textsc{sem} and the choice of the different parameters, e.g. Poisson
ratios, ...) with $t=0.67$ and $u=2.14$ when $\tau$ is set to $45^\circ$.
\label{sheardata}}
\efig

The pouring procedure involved a sieve which was slow\-ly raised in
order to create a uniform rain of sand. Thereafter the sand layer was
packed by pushing on the free-surface with a plate, and by
simultaneously taping on the lateral edges of the box. Note that these
conditions are similar to the so-called `dense preparation method'
proposed earlier by Serero \emph{et al.} \cite{SRCCL01}. A vertical
stress response was determined before and after shearing. During the
shear deformation of the sample, a weight ($\sim 40$kg) was imposed on
the top surface and the lateral boundaries (labeled as $2$ on the left
of figure \ref{fig:exp3D}) were slowly tilted up to a final angle
$\phi$. The dense preparation as well as the large imposed load seemed
necessary to avoid inhomogeneous deformations of the free-surface and
help to hinder the formation of localized shear-bands. This result was
confirmed by direct inspection of the displacement fields of the
grains lying on the transparent side of the lateral boundaries. In
order to monitor this displacement field, three vertical columns of
colored grains were inserted next to the lateral boundaries prior to
shearing (see right panel of figure \ref{fig:exp3D}).  We note that
the formation of a shear band was hindered only for limited values of
aspect ratios (typically for ratios $h/l$ from $0.3$ to $1$ used here)
and for shear angles smaller than $5^\circ$. We also point out that
due to the rather close-to-unit aspect ratio of the shear box, the
response for large horizontal distances between the source and the
stress probe was not measured so as to avoid the influence of lateral
walls.

In figure \ref{sheardata}, we show the experimental results for the
vertical stress response $\szz$ after shear for several shear angles
$\phi$ and layer depths $h$. In this figure, the abscisa is the
horizontal coordinate $x$ normalized by the sand depth $h$. We
normalize the stresses $\szz(x)$ by $\szz(0)$ the value of the stress
at $x=0$ (i.e. immediately below the piston). This normalization is
due to the finite lateral extension of the box since, contrary to
previous situations \cite{RC01,SRCCL01}, we cannot obtain a proper
normalization of the stress by integration (i.e. assess the total
force applied). For shear angles $\phi$ varying from $2^\circ$ to
$5^\circ$, the experiments clearly show, a skewing of the response
function in association with a displacement of the maximum
corresponding to an angle $\beta=8^\circ \pm 1^\circ$ (deduced from
the slope of the empirical relation between the height of sand and the
shift). Thus, as in the 2D case, the response indicates that, due to
shearing, the granular assembly is clearly anisotropic (at least, the
vertical direction is no longer an axis of symmetry). It is worth
mentioning that this tendency is robust, as it was observed for other
types of grains (larger, slightly polydisperse, rough sand grains of
diameter $\sim 1$mm, and smooth spherical beads of diameter $\sim
1.5$mm).

\bfig[t]
\bc
\epsfxsize=\linewidth
\epsfbox{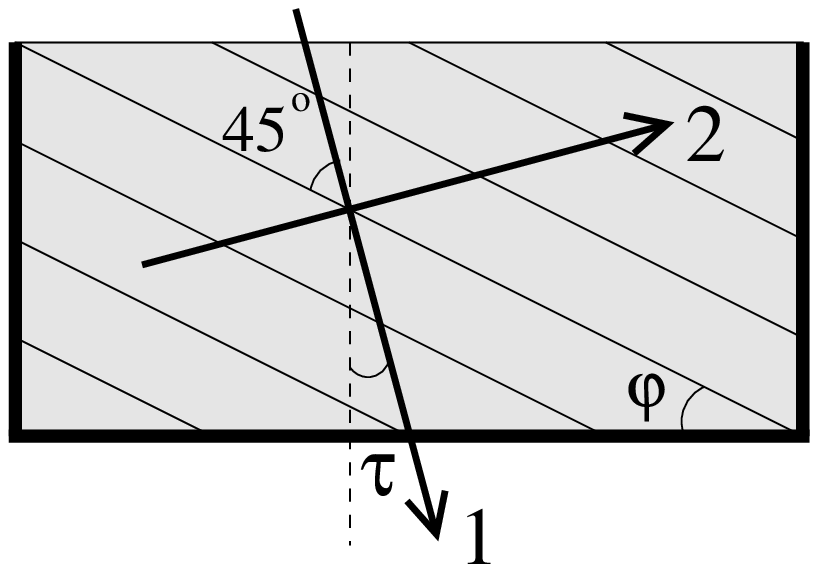}
\ec
\caption{Sketch of the avalanche preparation procedure. The avalanche angle
$\varphi=30^\circ$, so that the expected angle of anisotropy is $\tau=15^\circ$.
\label{avalanchebox}}
\efig
\bfig[t]
\bc
\epsfxsize=\linewidth
\epsfbox{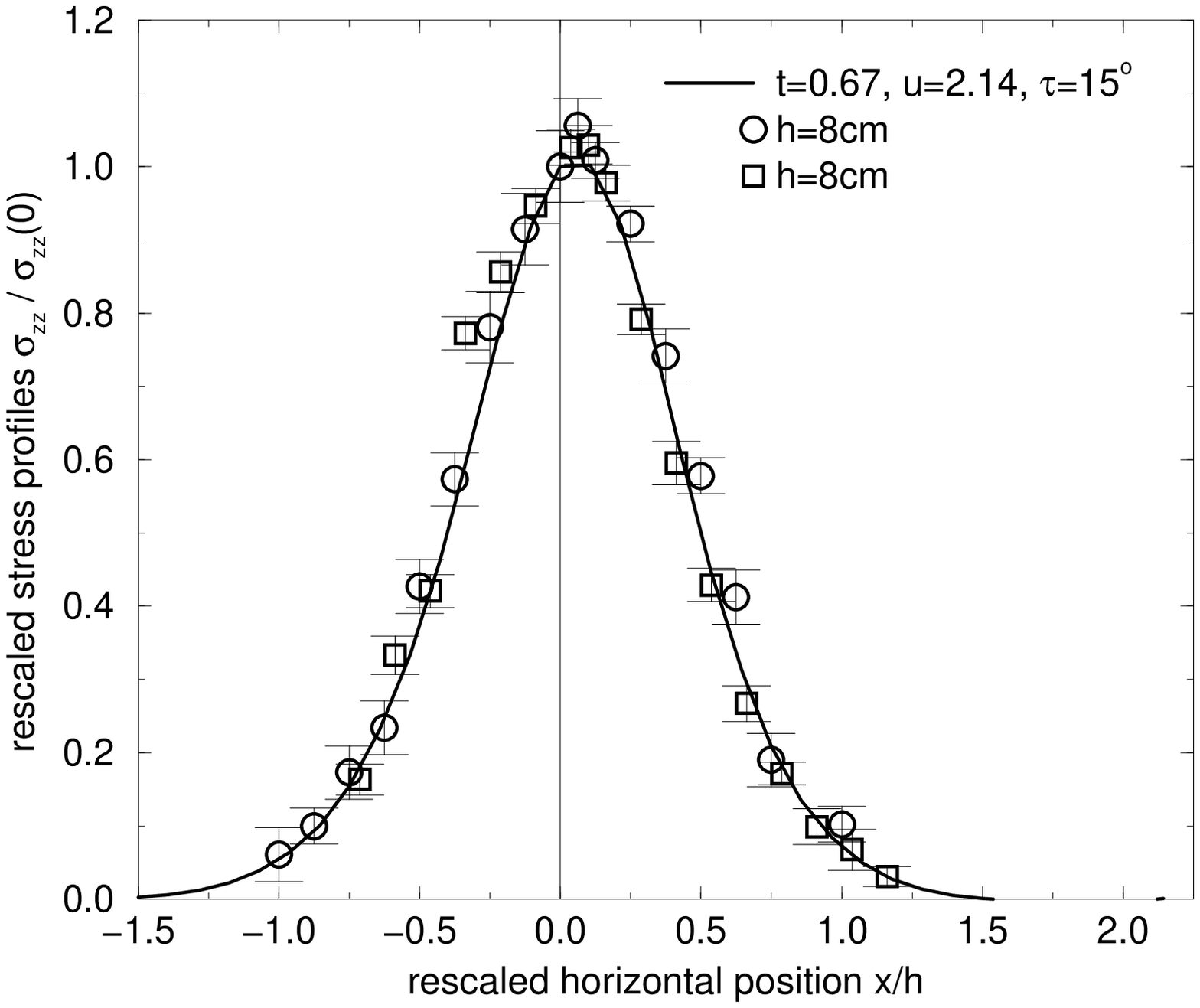}
\ec
\caption{Two response profiles of a 3D layer of Fontainebleau sand prepared by
successive avalanches. The thickness of the layer is $h=8$cm. In comparison
to figure \protect\ref{sheardata}, the shape of the profile is also well
reproduced by the \textsc{sem} (idem, see section \ref{secII}) with $t=0.67$ and
$u=2.14$ but now when $\tau$ is set to $15^\circ$.
\label{avaldata}}
\efig

\emph{The avalanche preparation} -- Next we prepared a layer of sand
structured by successive avalanches (see figure \ref{avalanchebox} for
a sketch of the preparation procedure). The experimental results for
two independent preparations are displayed in figure \ref{avaldata}
and rescaled in a way similar to the case of the 3D shear box. Again,
we observe a skewing of the response indicating an anisotropic texture
induced by the avalanches. However, the tilt effect of the response is
weaker, as it corresponds to an angle for the locus of the maxima
$\beta = 3.5^\circ \pm 1^\circ$.

%_____________________________________________________________________________
\section{Elastic calculations}
\label{secII}

We next turn to modeling the experimental results using elasticity
theory. We show here that a simple anisotropic description is
sufficient to capture all the salient features of the data. Isotropic
elasticity, in both 2D and 3D, has only two constitutive parameters
for a given material, namely the Young modulus $E$ and the Poisson
ratio $\nu$. When anisotropy is introduced, however, the number of
independent parameters increases significantly. For example, five
parameters are required for the case of simple 2D orthotropy that we
use below. In 3D, this number is even higher. Fortunately, for the
static equations, only combinations of these parameters enter
(e.g. the parameters $r$ and $t$, below). Still, a meaningful fit of
the response profiles to an anisotropic elastic model is a formidable
task. Although such an approach is very instructive, it is not an
essential point for the present purposes. We will present the results
of this type of approach elsewhere.

\subsection{Elastic response in 2D}

Let us first focus on the simpler two dimensional case. What is
indicated by the 2D shearing experiment described in the previous
section, is that the sheared force chain network clearly has a
preferred direction at an angle of $45^{o}$ with respect to the
initial axis of the box. Contacts between the grains are gained or
reinforced in this strong direction, whereas they are lost or weakened
in the perpendicular direction. It is then expected that an effective
elastic medium model for this situation would have a stiff direction
(direction 1) which makes a fixed angle $\tau =45^\circ$ relative to
the vertical axis $z$ ($x$ denotes the horizontal coordinate)
characterized by a Young modulus $E_1$.  Perpendicularly to direction
1 is the softer direction (direction 2) that is characterized by a
Young modulus $E_2<E_1$. This orthotropic elastic description is
completed by three additional parameters: two Poisson coefficients
$\nu_{12}$ and $\nu_{21}$ and the shear modulus $G$. Then, the
strain-stress relation, expressed in these tilted axis, can be written
in the following matrix form:
\be
\left ( \bay{c} \uuu \\ \udd \\ \uud \eay \right ) =
\left ( \bay{ccc}
\frac{1}{E_1} & -\frac{\nu_{21}}{E_2} & 0 \\
-\frac{\nu_{12}}{E_1} & \frac{1}{E_2} & 0 \\
0 & 0 & \frac{1}{2G}
\eay \right )
\!\!\! 
\left ( \bay{c} \suu \\ \sdd \\ \sud \eay \right ),
\label{strainstressequa}
\ee
Because the matrix involved in this relation must be symmetric, the Young's
moduli and Poisson ratios are not independent, but satisfy the relation
$E_2/E_1=\nu_{21}/\nu_{12}$. The elastic free energy is well defined if 
\beqa
E_1, E_2, G        & > & 0, \\
1-\nu_{12}\nu_{21} & > & 0.
\label{condstab2D}
\eeqa
The isotropic case is recovered for $E_1=E_2=E$, $\nu_{12}=\nu_{21}=\nu$
and $G=\frac{E}{2(1+\nu)}$.

In reference \cite{OBCS03}, Otto \emph{et al.} have analytically computed the
stress tensor components for such an anisotropic elastic material in the case of
a semi-infinite medium ($z>0$) loaded with a unitary point force at the origin 
$x=z=0$. They have shown that, for a given anisotropy angle $\tau$, the
stress response profiles only depend on two quantities that involve a
combination of the elastic coefficients:
\beqa
\label{deft}
t & = & E_2/E_1=\nu_{21}/\nu_{12},\\
\label{defr}
r & = & \frac{1}{2} \, E_2
\left ( \frac{1}{G} - \frac{\nu_{12}}{E_1}
                    - \frac{\nu_{21}}{E_2}
\right ).
\eeqa
It is notable that for other geometries, the stresses could
depend on the $E$'s, $\nu$'s or $G$ independently of $t$ and $r$ due to
the effect of different boundary conditions, for example at the
bottom of a finite thickness slab. By contrast, the
corresponding familiar isotropic solution is completely independent
of $E$ and $\nu$, although some (weak) variations of the stress
profiles with $\nu$ are found for an isotropic layer of finite
thickness \cite{SRCCL01,G73}. Depending on the values of $\tau$, and in
particular on the sign of $r$, and of $(r^{2}-t)$, different response
shapes result, including single or double peaked responses, and
symmetrical or skewed profiles (see numerous figures in \cite{OBCS03}).

\bfig[t]
\bc
\epsfxsize=\linewidth
\epsfbox{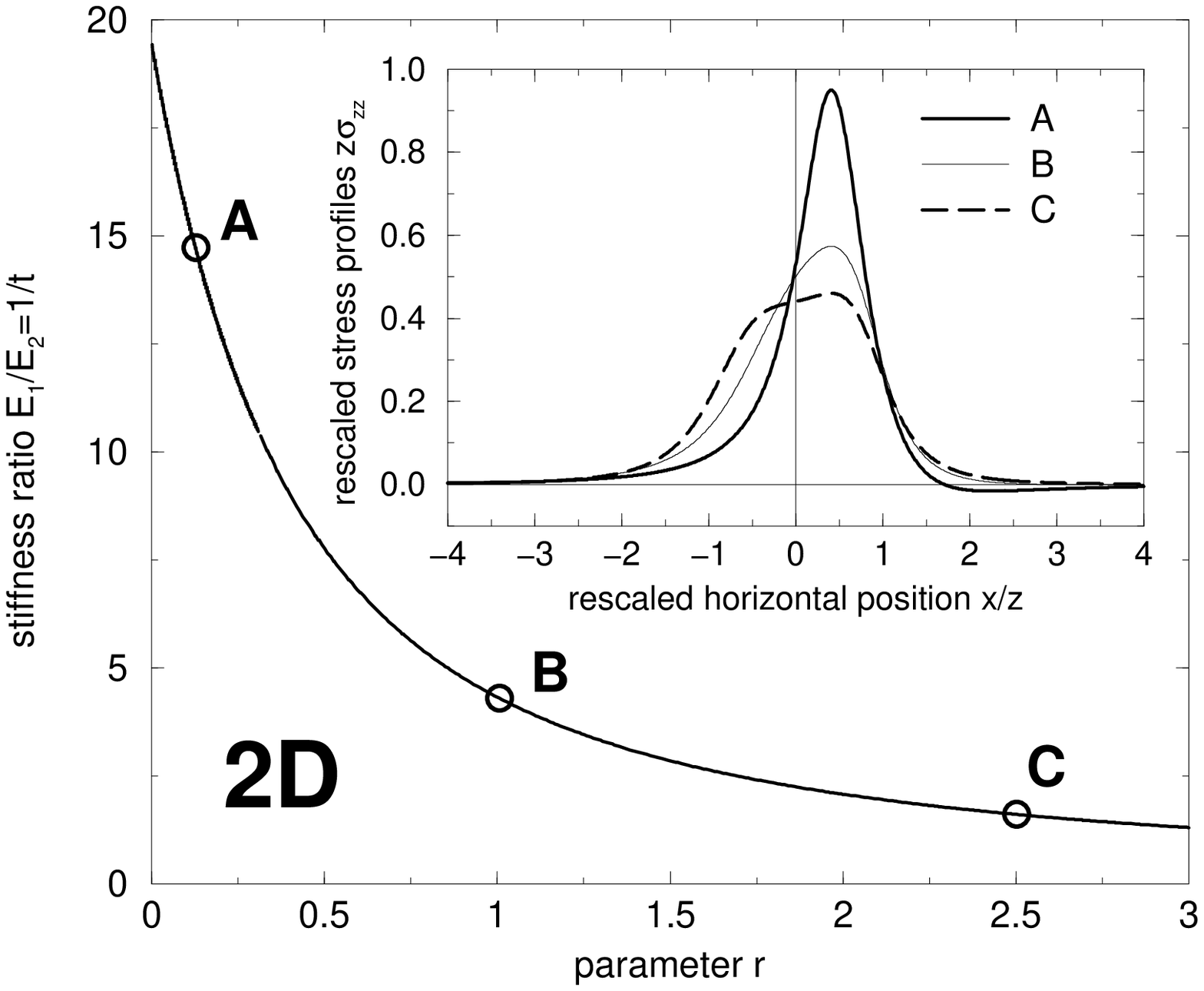}
\ec
\caption{Values of $r$ and $t$ which give a response profile for a
semi-infinite medium with a tilt angle of $\beta = 22^\circ$.
Inset: Three particular such stress profiles: $r=0.15$ and $t=0.07$ 
(A, bold line), $r=1.0$ and $t=0.24$ (B, thin line), and $r=2.5$ and $t=0.62$
(C, dashed line). The response shape is wider for larger $r$.
\label{beta=22}}
\efig

Although it is natural to fix the angle $\tau$ and to use the ratio
$E_2/E_1$ as a control parameter for the amplitude of the anisotropy
created by the shearing, it is rather difficult to have any kind of
intuition about the $\nu$'s and the shear modulus $G$, i.e. on
$r$. Figure \ref{beta=22} shows the locus of values of $r$ and $t$
which give a response profile for a semi-infinite medium with a
deflection angle of $\beta =22^\circ$, the experimental
value. Interestingly, we see the tendency for a higher value of $r$ to
significantly broaden the response function. This would correspond,
for example, to small values of the shear modulus $G$. Note also, that
the width at half-height of the response is a quantity that is easily
accessible experimentally, and hence provides a good method to test
this type of model.

\subsection{Elastic response in 3D}

It is not difficult to generalize this approach to 3D. However, in
this case, two axes (directions $2$ and $3$) orthogonal to direction 1
must be specified. We keep the notation $x$ for the horizontal axis
along which the shearing is applied, and the stress response
measured. Direction $2$ is in the vertical plane $(z,x)$, and
direction $3$ $(y)$ is perpendicular to these two directions.

The matrix involved in the 3D equivalent of the strain-stress relations
(\ref{strainstressequa}) has a similar structure to the 2D case, but with
three Young moduli $E_{i}$, three shear moduli $G_{i}$ and six Poisson
ratios $\nu_{ij}$ ($i$ and $j$ must be different). Again, the symmetry of
the matrix gives three relations of the form $\nu_{ij}/E_{i}=\nu_{ji}/E_{j}$.
In the final analysis, there are $9$ independent coefficients besides the
anisotropy angle $\tau$. Stability now requires that 
\beqa
E_i, G_i & > & 0, \\
1-\nu_{ij}\nu_{ji} & > & 0, \\
1 - \sum_{i \neq j} \nu_{ij}\nu_{ji}
  - \nu_{12}\nu_{23}\nu_{31} - \nu_{21}\nu_{32}\nu_{13} & > & 0.
\label{condstab3D}
\eeqa
Unfortunately, in 3D, we do not yet have closed form analytic expressions
for the stress response profiles comparable to those given in \cite{OBCS03}.
To compute profiles in 3D, we make use of a finite element free-ware code
developed in soil mechanics called \textsc{castem} \cite{Castem}. We typically
chose a grid with $40 \times 40 \times 12$ cells in the $x$, $y$ and $z$ direction
respectively. In this subsection, we show that the trends suggested by the
experimental data are reproduced rather well in a model that allows for a
modification of the constitutive elastic relations with respect to the
symmetry of the external deformation applied to the system.

\bfig[t]
\bc
\epsfxsize=\linewidth
\epsfbox{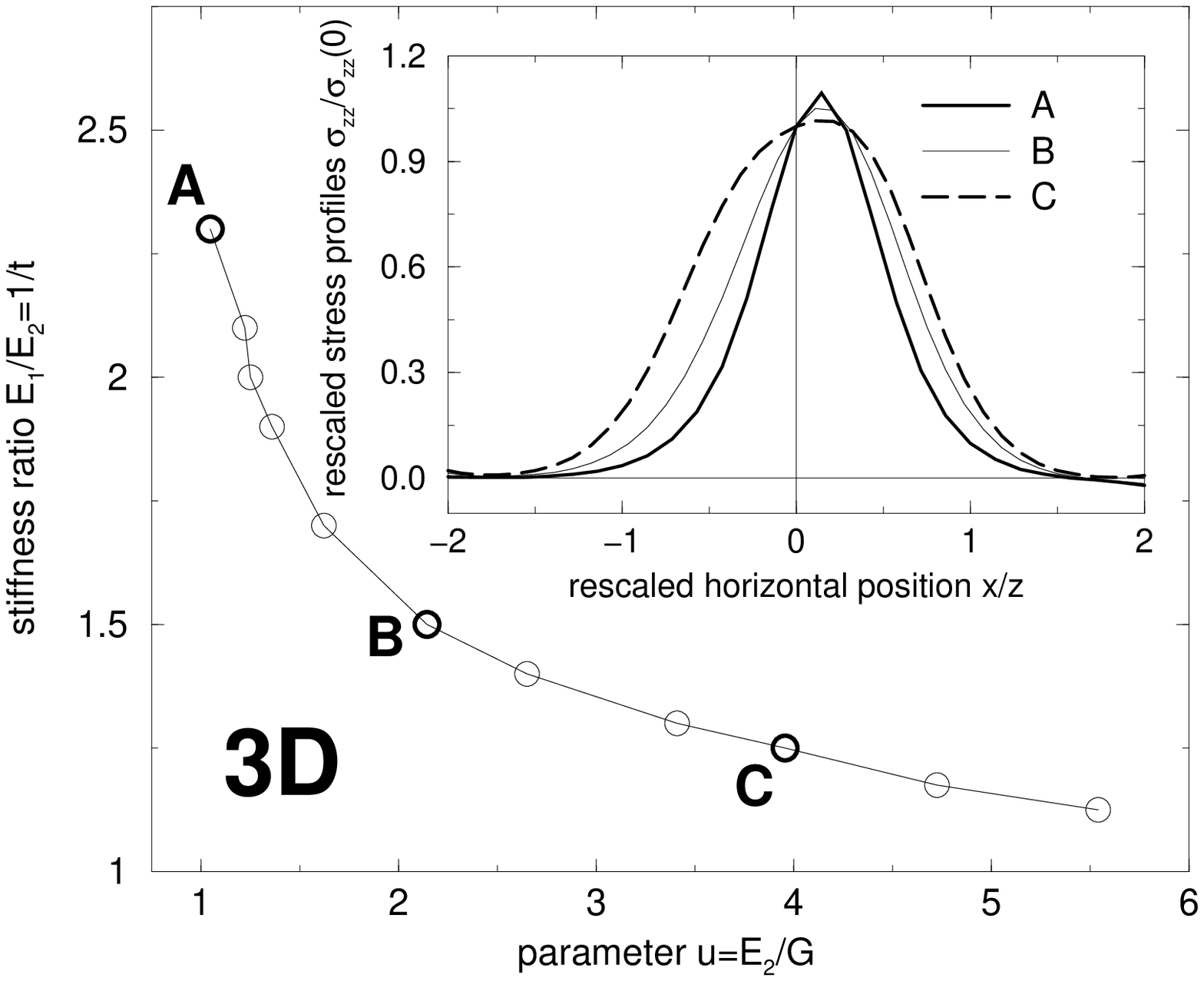}
\ec
\caption{3D equivalent of figure \protect\ref{beta=22}:
values of $u$ and $t$ which give a response profile
with a tilt angle of $\beta = 8^\circ$. Note however that these 3D
calculations have been performed numerically with \textsc{castem}
on a layer of \emph{finite} thickness.
Inset: Three particular such stress profiles: $u=1.05$ and $t=0.43$
(A, bold line), $u=2.14$ and $t=0.67$ (B, thin line), and $u=3.96$ and
$t=0.8$ (C, dashed line). The response shape is wider for larger $u$.
\label{beta=8}}
\efig

Here, we analyze the previous 3D experiments on the shear box and on
the avalanche preparation. We emphasize that we do not make a detailed
fit of the experimental data to a general anisotropic elasticity
model. Using the numerical code, we only explore the possibilities of
orthotropic elasticity on a `Simple Elastic Model' that we call
\textsc{sem} in the sequel. Again, we do not seek a complete
exploration of the parameter space and thus, we only vary three
parameters that we believe are crucial namely, the stiffness of
the direction of anisotropy $E_1$, the shear modulus $G$ and $\tau$ the
direction of the orthotropic axis with respect to the vertical
direction. More precisely, we consider the same value of $G$ for all
directions, i.e.  $G_{1}=G_{2}=G_{3}=G$. The average Young's modulus is
taken constant: $E=(E_1+E_2+E_3)/3=150$MPa. We also take $E_2=E_3$.
Finally, the three
Poisson coefficients $\nu_{12}$, $\nu_{13}$ and $\nu_{23}$ are also
kept constant and bear the values $\nu =0.3$. We checked that the value
of this parameter is not very sensitive. The remaining $\nu_{ij}$
are such that the strain-stress matrix is symmetrical (see above). In
the spirit of the 2D analytical results, we use two dimensionless
parameters to present our data, i.e. the stiffness ratio
$t=E_{2}/E_{1}$, and the shear ratio $u=E_{2}/G$. $u$ is analogous to
the parameter $r$ in 2D -- see equation (\ref{defr}).

\emph{The shear cell} -- An essential result of the shear box
experiment was that the tilt angle of the stress response profiles is
around $8^\circ$ (figure \ref{sheardata}).  As suggested in the 2D space,
it is natural that the orthotropic direction is tilted with an angle $\tau
=45^\circ$. In figure \ref{beta=8} we thus present the curve $t(u)$
corresponding to a tilt angle $\beta =8^\circ$ and for $\tau=45^\circ$
as computed by \textsc{castem} using the \textsc{sem}. We see that
similarly to the 2D situation, moving on this curve from lower to
higher values of $u$ significantly enlarges the width of the response
function (see inset of figure \ref{beta=8}). This is indeed an
experimental observable that can be used to discriminate in parameter
space. In the framework of the \textsc{sem}, good agreement with the
experimental data on the shear box can be obtained for values
$u_0=2.14$ and $t_0=0.67$. The curve corresponding to $(u_0,t_0)$ is
presented in figure \ref{sheardata}.

\emph{The avalanche preparation} -- Now, we seek an interpretation of
the response function for the granular slab prepared by successive
avalanches. First we note that during the avalanching process, the
flowing layer experiences a strong shear along the avalanche angle
$\varphi$. An interesting outcome of the shear box experiment is that
for all finite shear angle values $\phi$ that we tested, the shape of
the response function as well as tilt angle $\beta$ did not vary
significantly. We argue that each deposited layer retains a memory of
the shear due to the avalanching process, which induces an anisotropic
texture. In first approximation, we may assume that, the avalanche
acts like a shear box whose bottom is tilted at a angle $\varphi \sim
30^\circ$ with respect to the horizontal direction. We propose to use
the parameters $(u_0,t_0)$ which compares reasonably to the shear box
data to interpretate the avalanche deposition experiment. Of course
now, we need to tilt the orthotropic angle $\tau$ from $45^\circ$ to a
value $\tau =45^\circ - \varphi \sim 15^\circ$. In figure
\ref{avaldata} we present the results of \textsc{castem} calculations
on the \textsc{sem} using these parameters and indeed we find an angle
$\beta =3.0^\circ \pm 0.5^\circ$, close to the experimental value
$\beta =3.5^\circ \pm 1^\circ$.

%_____________________________________________________________________________
\section{Back to the sandpile pressure profile}
\label{secIII}

Several experimental determinations of the stress distribution below a
sand pile prepared from a point source were obtained under various
experimental conditions. First, we note that all stress data can be
rescaled so that it is possible to compare the prediction of the
elastic model with experiments made with different pile sizes. For a
conical pile of height $h$, radius at the base $R=h/\tan\varphi$, and
pile slope angle $\varphi$, the relevant parameters are the rescaled
vertical stress $\szz/\rho gh$ and the rescaled horizontal position
$\xi =r/R$ ($\rho$ is the density of the granular packing and $g$ the
gravity constant).

\bfig[t]
\bc
\epsfxsize=\linewidth
\epsfbox{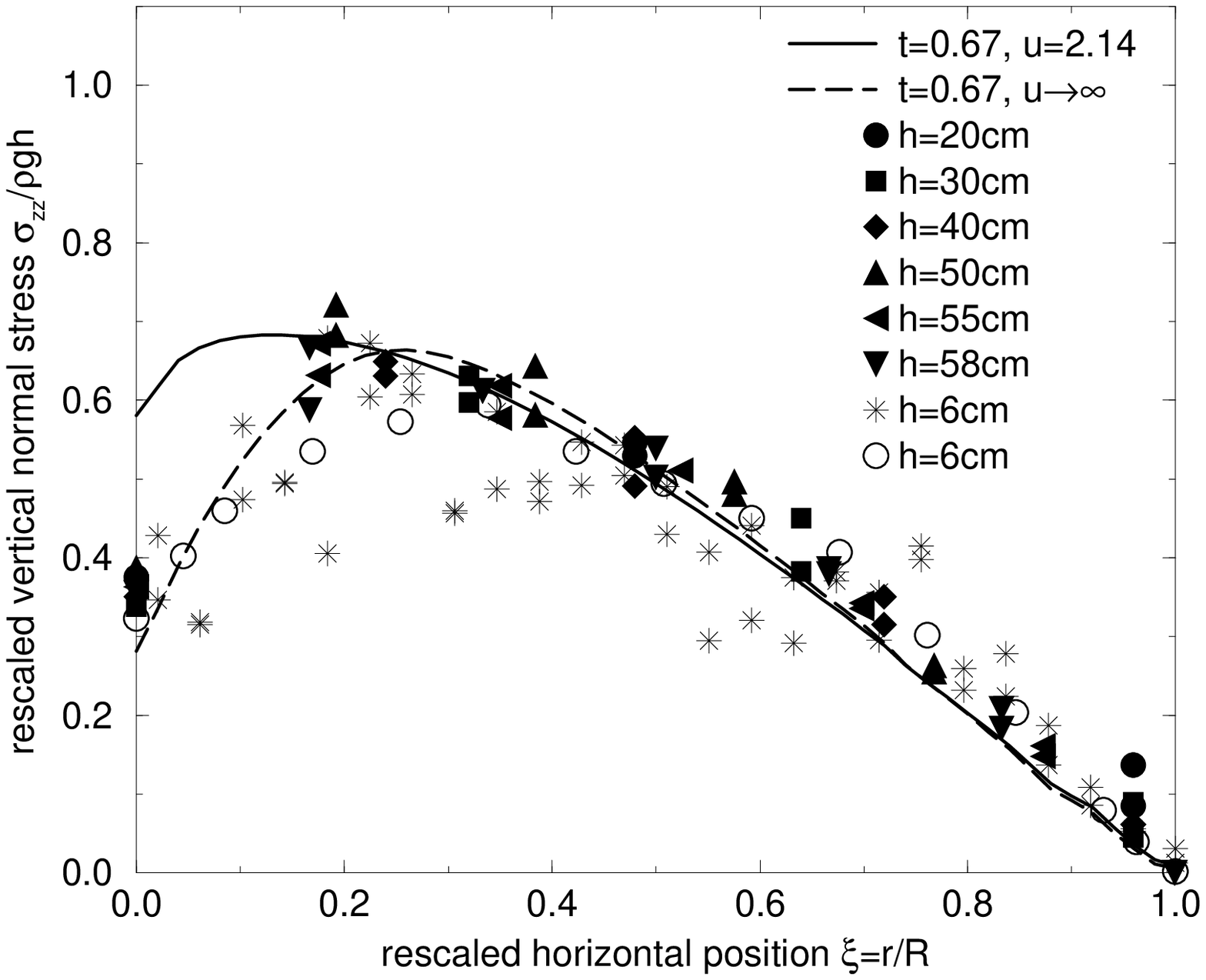}
\ec
\caption{Normal stress below a conical pile. Filled symbols are the data
of \v{S}m\'{i}d and Novosad \cite{SN81}, the stars are from Brockbank
\emph{et al.} \cite{BHB97}, and the empty circles show Vanel
\emph{et al.} measurements \cite{VHCBC99}. The corresponding pile heights $h$
are indicated in legend. The repose angles of the material used by these authors
are respectively $\varphi=33^\circ$, $\varphi=31^\circ$ and $\varphi=30^\circ$.
For simplicity we have set $\varphi=30^\circ$ in all our calculations.
The solid line is the prediction of the \textsc{sem} in the conical geometry
with $t=0.67$ and $u=2.14$ when $\tau$ is set to $15^\circ$. The dashed line
is the same except for $u\to\infty$, which seems to fit better the data.
\label{dipdata}}
\efig

In figure \ref{dipdata}, we show the experimental data obtained by
three groups, \v{S}m\'{i}d and Novosad \cite{SN81}, Brockbank \emph{et
al.}  \cite{BHB97} and Vanel \emph{et al.} \cite{VHCBC99}.  The
underlying idea behind the use of an anisotropic elasticity is to
include the frozen texture caused by shearing that ensues during
avalanche deposition. From the response function experiments under
shearing (shear box and avalanche) displayed before, it is natural to
propose a modelling of the sand pile prepared by successive avalanche
using an orthotropic elastic model like \textsc{sem}. Of course, we
see here the limitations of such a modelling, since in the
\textsc{sem} version of the \textsc{castem} computation we only have
three parameters to vary. Furthermore, as discussed previously, in the
case where the sand pile repose angle is $\varphi=30^\circ$, the
orthotropic direction should be fixed at $\tau =15^\circ$. We show in
figure \ref{dipdata} the results of a \textsc{castem} computation for
a conical sand pile using the `optimal' parameters $(u_0,t_0)$ that
yielded a reasonable agreement to the shear box and the avalanche
deposition experiments. Note that we also checked that the stress state
is always below the Coulomb criterion (with an internal friction angle of
$30^\circ$), i.e. that failure never occurs -- except marginally at the
very free surface as all stress components vanish.

We observe the presence of a dip that bears essentially the correct
qualitative features. Nevertheless, on a quantitative level, the
values of the depth of the dip as well as the amplitude of the stress
maximum seem too large. Furthermore, the value of $\xi _{max} \sim
0.15$ underestimates clearly the experimental data. In fact the data
seem to be better fitted by the \textsc{sem} with a very small shear
modulus, i.e. $u\to\infty$. The understanding of this observation remains
an open question.

%_____________________________________________________________________________
\section{Conclusion}

In this paper we present an analysis of several response function
experiments under the scope of anisotropic elasticity. As an
application, we investigate to which extent the pressure dip observed
at the bottom of a sand pile prepared by successive avalanches can be
understood as a texturing effect induced by shearing along the
avalanche directions.

In the context of an elastic analysis, the orthotropic model used in
this paper is relatively simple, since its main features are the
presence of a stiff axis in the orthotropic direction and a shear
modulus that can be varied independently. The orthotropy axis
direction is a third parameter simply taken at an angle $\pi /4$ with
respect to
the shearing direction. This choice corresponds to the direction of
principal compression and is consistent with our recent experimental
findings in 2D \cite{GRCB03}. We also present new sets of experiments
on a 3D sand packing in a shear box, where shear induced anisotropy is
present. The second experiment is a sheared granular slab constructed
by successive avalanches. The agreement with the orthotropic model is
quantitatively correct, as the tilt of the response function can be
reproduced by elastic modelling in both experiments using the same set
of parameters.

In a second series of analysis, we use the best fit parameters
obtained from comparison with response function experiments to see
whether such a relatively simple modelling is likely to explain the
dip below a sand pile, as suggested initially by Savage
\cite{S98}. Indeed from a finite element calculation of a conical pile,
we obtain a good qualitative agreement with the available
experimental data but we fail to obtain really quantitative
agreement, as the dip amplitude and its width seem both
underestimated by a factor of almost $30\%$ in the central part of
the pile. It is not yet clear whether the elasticity model we use is
oversimplified since it contains only two parameters which are
sensitive to shear, or whether the preparation procedure under
avalanches is not accounted for correctly in the context of the 3D
sand-pile. We note here the interesting suggestion of Jenkins
\cite{J} that the upward moving `stoppage' waves produced when the
avalanche hits the ground could modify the main compression axis so
that its direction could be further away from the vertical. This effect
would indeed enlarge the size of the dip.

The tendency for deposition history or external action like shear or a
biaxial compression to modify the constitutive structure of a material
was noted in experiments by Oda \emph{ et al.} \cite{ONK85} and in
numerical simulations by Radjai \emph{et al.} \cite{RWJM98} and is at
present time still a very open and difficult question. This paper
calls for more extensive systematic studies both experimentally and
numerically (possibly analytically). The sand pile pressure dip, as
interesting as it seems, appears a bit too complex to analyse for the
moment. For a better understanding of this crucial issue, this work
suggests a systematic use of response function techniques -- stress
responses as here, as well as displacement responses as in
\cite{KCJCC04,LTWB04,MPRV04} -- so as
to extract several effective constitutive parameters of the material
along a given stress-strain history. This is a promising systematic
approach, and a possible alternative to sound propagation techniques,
which can precisely identify internal structural changes due to
external action a granular material \cite{ABGRCCBC04}. This is the
scope of ongoing projects in our laboratories.

%_____________________________________________________________________________

\noindent
\rule[0.1cm]{3cm}{1pt}

We thank C.~Goldenberg, I.~Goldhirsch, J.~Jenkins, J. Lanuza, S.~Luding and
J.~Snoeijer for fruitful discussions.
A.P.F. Atman's present address is Departamento de F\'\i sica, Instituto de
Ci\^encias Exatas, Universidade Federal de Minas Gerais, C.P. 702, 30123-970,
Belo Horizonte, MG, Brazil.
P. Brunet's present address is Royal Institute of Technology - Department of
Mechanics. Teknikringen 8, 10044 Stockholm, Sweden.
The LPMMH is UMR 7636 of the CNRS.

%_____________________________________________________________________________


\begin{thebibliography}{99}

\bibitem{reviews}
For a broad perspective on granular materials, see
The focus issue on the physics of granular media of the Comptes-Rendus
de l'Acad\'emie des Sciences, Physique \textbf{3}, pp 129-245 (2002);
the focus issue on granular materials of Chaos \textbf{9}, pp 509-696 (1999);
\textit{Physics of Dry Granular Media}, H.J. Herrmann,
J.-P. Hovi, and S. Luding editors, NATO ASI Series, Kluwer, (1997);
\textit{Powders and Grains 97},
R.P. Behringer and J.T. Jenkins editors, Balkema, (1997);
H.M. Jaeger, S.R. Nagel, and R.P. Behringer,
Rev. Mod. Phys. \textbf{68}, 1259 (1996).

\bibitem{SN81}
J. \v{S}m\'{\i}d, and J. Novosad,
Proc. Powtech. Conference
1981, Ind. Chem. Eng. Symp. \textbf{63}, D3V 1 (1981).

\bibitem{BHB97}
R. Brockbank , J.M. Huntley, and R.C. Ball,
J. Phys. (France) II \textbf{7}, 1521 (1997).

\bibitem{VHCBC99}
L. Vanel, D.W. Howell, D. Clark, R.P. Behringer and E. Cl\'ement,
Phys. Rev. E \textbf{60}, R5040 (1999).

\bibitem{GLBH01}
J. Geng, E. Longhi, R.P. Behringer and D.W. Howell,
Phys. Rev. E \textbf{64}, 060301(R) (2001).

\bibitem{LCH92}
K. Liffman, D.Y.C. Chan and B.D. Hughes,
Powder Technology \textbf{72}, 255 (1992).

\bibitem{L97}
S. Luding,
Phys. Rev. E \textbf{55}, 4720 (1997).

\bibitem{M98}
H.-G. Matuttis,
Granular Matter \textbf{1}, 83 (1998).

\bibitem{BCC95}
J.-P. Bouchaud, M.E. Cates and P. Claudin,
J. Phys. (France) I \textbf{5}, 639 (1995).

\bibitem{WCCB96}
J.P. Wittmer, P. Claudin, M.E. Cates and J.-P. Bouchaud,
Nature, \textbf{382}, 336 (1996).

\bibitem{WCC97}
J.P. Wittmer, M.E. Cates and P. Claudin,
J. Phys. (France) I \textbf{7}, 39 (1997).

\bibitem{C99}
P. Claudin,
Ph.D. thesis \textit{La physique des tas de sable},
Annales de Physique \textbf{24}, n$^\circ$2, 1 (1999).

\bibitem{VCBCC00}
L. Vanel, P. Claudin, J.-P. Bouchaud, M.E. Cates, E. Cl\'ement and J.P. Wittmer,
Phys. Rev. Lett. \textbf{84}, 1439 (2000).

\bibitem{YY}
J.-P. Bouchaud, P. Claudin, D. Levine and M. Otto,
Eur. Phys. J. E \textbf{4}, 451 (2001);
J.E.S. Socolar, D.G. Schaeffer and P. Claudin,
Eur. Phys. J. E \textbf{7}, 353 (2002)
+ erratum EPJE \textbf{8}, 453 (2002);
Y. Roichman, D. Levine and I. Yavneh,
Phys. Rev. E \textbf{70}, 061301 (2004).

\bibitem{B04}
R. Blumenfeld,
Phys. Rev. Lett. \textbf{93}, 108301 (2004).

\bibitem{CWBC98}
M.E. Cates, J.P. Wittmer, J.-P. Bouchaud and P. Claudin,
Phys. Rev. Lett. \textbf{81}, 1841 (1998);
Phil. Trans. R. Soc. London A \textbf{356}, pp 2535-2560 (1998);
Chaos \textbf{9}, 511 (1999).

\bibitem{W90}
D.M. Wood,
\textit{Soil Behaviour and Critical State Soil Mechanics},
Cambridge University Press, Cambridge (1990).

\bibitem{MGJS99}
H. Makse, N. Gland, D.L. Johnson and L.M. Schwartz,
Phys. Rev. Lett. {\bf 83}, 50705073 (1999).

\bibitem{dG98}
P.-G. de Gennes,
Physica A \textbf{261}, 267 (1998);
Rev. Mod. Phys. \textbf{71}, S374 (1999).

\bibitem{SR00}
M. da Silva and J. Rajchenbach,
Nature \textbf{406}, 708 (2000).

\bibitem{RC01}
G. Reydellet and E. Cl\'ement,
Phys. Rev. Lett. \textbf{86}, 3308 (2001)

\bibitem{SRCCL01}
D. Serero, G. Reydellet, P. Claudin, E. Cl\'ement and D. Levine,
Eur. Phys. J. E \textbf{6}, 169 (2001).

\bibitem{GHLBRV01}
J. Geng, D. Howell, E. Longhi, R.P. Behringer, G. Reydellet, L. Vanel,
E. Cl\'ement and S. Luding,
Phys. Rev. Lett. \textbf{87}, 035506 (2001).

\bibitem{MJN02}
N.W. Mueggenburg, H.M. Jaeger and S.R. Nagel,
Phys. Rev. E \textbf{66}, 031304 (2002).

\bibitem{GRCB03}
J. Geng, G. Reydellet, E. Cl\'ement and R.P. Behringer, 
Physica D \textbf{182}, 274 (2003).

\bibitem{SMJN03}
M.J. Spannuth, N.W. Mueggenburg, H.M. Jaeger and S.R. Nagel,
\texttt{cond-mat/0308580}.

\bibitem{EC97}
C. Eloy and E. Cl\'ement,
J. Phys. I \textbf{7}, 1541 (1997).

\bibitem{M00}
J.-J. Moreau,
in the proceedings of the
\textit{colloque `Physique et m\'ecanique des mat\'eriaux granulaires'},
Champs-sur-Marne (France), 199 (2000).

\bibitem{BCCZ02}
L. Breton, P. Claudin, E. Cl\'ement and J.-D. Zucker,
Europhys. Lett. \textbf{60}, 813 (2002).

\bibitem{GG02a}
C. Goldenberg and I. Goldhirsch,
Phys. Rev. Lett. \textbf{89}, 084302 (2002).

\bibitem{GG04}
C. Goldenberg and I. Goldhirsch,
Granular Matter \textbf{6}, 87 (2004).

\bibitem{SVG02}
R. da Silveira, G. Vidalenc and C. Gay,
\texttt{cond-mat/0208214}.

\bibitem{OP04}
S. Ostojic and D. Panja,
\texttt{cond-mat/0403321}, \texttt{cond-mat/0409160}.

\bibitem{ABGRCCBC04}
A.P.F. Atman, P. Brunet, J. Geng, G. Reydellet, G. Combe, P. Claudin,
R.P. Behringer and E. Cl\'ement,
to appear in J. Phys. Cond. Mat. special issue on Granular Materials
(M. Nicodemi Editor), \texttt{cond-mat/0411734}.

\bibitem{GWM04}
N. Gland, P. Wang and H.A. Makse,
preprint (2004).

\bibitem{GG05}
C. Goldenberg and I. Goldhirsch,
to appear in Nature (2005).

\bibitem{GG02}
C. Goldenberg and I. Goldhirsch,
Eur. Phys. J. E \textbf{9}, 245 (2002).

\bibitem{OBCS03}
M. Otto, J.-P. Bouchaud, P. Claudin and J.E.S. Socolar,
Phys. Rev. E \textbf{67}, 031302 (2003).

\bibitem{HTW01}
D.A. Head, A.V. Tkachenko and T.A. Witten,
Eur. Phys. J. E \textbf{6}, 99 (2001);
see also the comment of J.-N. Roux, Eur. Phys. J. E \textbf{7}, 297 (2002).

\bibitem{KN04}
A. Kasahara and H. Nakanishi,
\texttt{cond-mat/0405169}.

\bibitem{EG99}
S.F. Edwards and D.V. Grinev,
Phys. Rev. Lett. \textbf{82}, 5397 (1999).

\bibitem{TW99}
A.V. Tkachenko and T.A. Witten,
Phys. Rev. E \textbf{60}, 687 (1999).

\bibitem{BB02}
R.C. Ball and R. Blumenfeld,
Phys. Rev. Lett. \textbf{88}, 115505 (2002).

\bibitem{M02}
C.F. Moukarzel,
J. Phys. Condens. Matter \textbf{14}, 2379 (2002).

\bibitem{R00}
J.-N. Roux,
Phys. Rev. E \textbf{61}, 6802 (2000).

\bibitem{S98}
S.B. Savage,
in `Physics of Dry Granular Media', H.J. Herrmann, J.P. Hovi and
S. Luding editors, NATO ASI series, Kluver Amsterdam, 25 (1998).

\bibitem{DCG00}
A.K. Didwania, F. Cantelaube and J.D. Goddard,
Proc. R. Soc. Lond. A \textbf{456} 2569 (2000).

\bibitem{G73}
J. Garnier,
\textit{Tassement et contraintes. Influence de la rigidit\'e de la fondation
et de l'anisotropie du massif.},
PhD thesis, Universit\'e de Grenoble (1973).

\bibitem{Castem}
with CAST3M, see \texttt{http://www.castem.org:8001}.

\bibitem{J}
J. Jenkins, private comm.

\bibitem{ONK85}
M. Oda, S. Nemat-Nasser and J. Konishi, 
Oils and Fundations \textbf{25}, 85 (1985).

\bibitem{RWJM98}
F. Radjai, D. Wolf, M. Jean and J.-J. Moreau,
Phys. Rev. Lett. \textbf{80}, 61 (1998).

\bibitem{KCJCC04}
E. Kolb, J. Cviklinski, J. Lanuza, P. Claudin and E. Cl\'ement,
Phys. Rev. E \textbf{69}, 031306 (2004).

\bibitem{LTWB04}
F. Leonforte, A. Tanguy, J.P. Wittmer and J.-L. Barrat,
Phys. Rev. B \textbf{70}, 014203 (2004).

\bibitem{MPRV04}
C.F. Moukarzel, H. Pacheco-Martinaez, J.C. Ruiz-Suarez and A.M. Vidales,
Granular matter \textbf{6}, 61 (2004).


\end{thebibliography}
\end{document}